\documentclass[twocolumn,showpacs,preprintnumbers,amsmath,amssymb,epsf]{revtex4-1}

\usepackage{graphicx,epsfig}
\usepackage{dcolumn}
\usepackage{bm}
\newcommand{\be}{\begin{equation}}
\newcommand{\ee}{\end{equation}}
\newcommand{\bse}{\begin{subequations}}
\newcommand{\ese}{\end{subequations}}
\newcommand{\bary}{\begin{eqnarray}}
\newcommand{\eary}{\end{eqnarray}}
\newcommand{\bwt}{\begin{widetext}}
\newcommand{\ewt}{\end{widetext}}

\begin{document}


\title{Hadronic-origin TeV flare of M87 in
  April 2010}
\author{Sarira Sahu$^{1}$,  Eddie Palacios$^{2}$
}
\affiliation{
$^{1}$Instituto de Ciencias Nucleares, Universidad Nacional Aut\'onoma de M\'exico, 
Circuito Exterior, C.U., A. Postal 70-543, 04510 Mexico DF, Mexico\\
$^{2}$ Facultad de F\'isica e Inteligencia Artificial, Universidad
Veracruzana, Sebasti\'an Camacho \# 5 , C. P. 91000,
Xalapa, Veracruz, Mexico
}


\begin{abstract}

M87 is a giant radio galaxy with FR-I morphology. It underwent
three episodes of TeV flaring in recent years with the strongest one in
April 2010 which was jointly monitored by MAGIC, VERITAS and
H.E.S.S. We explain  its spectral energy distribution in the energy range
0.3 TeV to 5 TeV by assuming that the flaring occurs in the innermost
region of the jet. In this region the low energy SSC photons serve as the
target for the Fermi-accelerated high energy protons of energy $\lesssim 30$
TeV to form a delta resonance. The TeV photons are produced from the
subsequent decay of the delta resonance to neutral pions.
In this scenario the observed TeV flux of the 2010 flare is fitted
very well.

\end{abstract}

\maketitle

\section{Introduction}

M87 is a giant radio galaxy in the Virgo cluster  
at a luminosity distance of $16.7\pm0.2$ Mpc\cite{Mei:2007xs} and a redshift of
$z=0.00436$.  The mass of the central supermassive black hole (SMBH)
is estimated be 
$M_{BH} = (3-6)\times10^{9} M_{\odot}$\cite{Macchetto:1997gi}. Based on its radio morphology
it is 
classified as FR-I galaxy\cite{Biretta:1999ja}. The radio images and modeling of its 
interaction with the surrounding environment suggests that the jet is
misaligned 
with respect to the line of sight\cite{Biretta:1995ja,Biretta:1999ja}. 
The substructures in the plasma jet originated from the center of M87 is resolved 
at different wavelengths (radio\cite{Cheung:2007wp}, optical\cite{Biretta:1999ja} and x-ray\cite{Harris:2011zf}). Due to the 
harboring of SMBH in the center and the presence of the jet, M87 was 
considered as a potencial candidate of TeV-emission. The evidence 
for very high energy (VHE) $\gamma$-rays ($E_{\gamma}>100$ GeV) emission 
from M87 was reported by the HEGRA Collaboration in 2003\cite{Reimer:2008a} and was 
later confirmed by H.E.S.S., VERITAS\cite{Aliu:2011xm,Abramowski:2011ze} and MAGIC. The AGN M87 is normally a weak 
VHE source, but this source shows strong variability at VHE with time scales 
of the order of days, which indicates a compact emission region 
$< 5\times10^{15}{\cal D}$ cm, (where ${\cal D}$ is the Doppler factor of 
the emitting plasma), corresponding to only a few Schwarzschild radii 
$R_{s}=2GM_{BH}/c^{2}$ $\simeq 10^{15}cm$.

So far, there are three episodes of enhanced VHE $\gamma$-ray emission 
observed from the AGN M87 in the years
2005\cite{Aharonian:2006ys,Abdo:2009ta}, 2008\cite{Aliu:2011xm} 
and 2010\cite{Aliu:2011xm,Ong:2010t}. The latest one of  April 2010, is the strongest TeV $\gamma$-ray flare ever 
detected from the AGN M87 with a peak flux of $(2.7\pm 0.68)\times
10^{-11}\,{\rm cm^{-2}\,{\rm s}^{-1}}$ for $E_{\gamma} > 350$ GeV\cite{Raue:2011zg,Abramowski:2011ze,Aliu:2011xm}. The
detected single isolated flare is well described by two sided
exponential functions with significantly different flux rise and decay
times\cite{Aliu:2011xm}. The rising (5 to 8 of April), peak (9 to 10 of April) and
falling (11 to 15 of April) parts of the flux during the flare
are consistent with  power-law behavior.
This flare was detected simultaneously by 
VERITAS, MAGIC and H.E.S.S.\cite{Ong:2010t,Abramowski:2011ze}, and triggered further multi-wavelength
observations in radio, optical and x-ray. This was also observed by
{\it Fermi}-LAT at MeV-GeV energies but could not observe day-scale
variability\cite{Raue:2011zg}.

Different theoretical models have been proposed to explain the flaring
in M87. Wagner et al.\cite{Wagner:2009zz} have complied the multi-wavelength data sets
spanning almost all the energy range and presented a spectral energy
distribution (SED) of M87 along
with leptonic and hadronic models predictions. The hadronic
synchrotron-proton blazar model\cite{Reimer:2004a} suggests  emission of synchrotron
photons from protons, charged pions and muons in the jet magnetic
field. However the SED produced using the archival data before 2004
shows a steep drop-off at TeV energies and to explain above TeV energy
 a strong Doppler boosting in needed which is not the case in M87. So
 this model is not compatible with any of the VHE spectral
 measurements after 2004. 
The leptonic decelerating inner jet model by Georganopoulos et al.\cite{Georganopoulos:2005yx} 
does not describe the hard TeV spectra well as it has a strong
cut-off. The multi-blob synchrotron self Compton (SSC) model by Lenain
et al.\cite{Reimer:2008a} needs a low magnetic field in the VHE emitting region which is
unlikely because this region is of the order of the Schwarzschild
radius and is expected to have strong field. Thus the so called one-zone
homogeneous leptonic models of  Georganopoulos et al.\cite{Georganopoulos:2005yx}  and Lenain
et al.\cite{Reimer:2008a} are very unlikely to reproduce the observed
VHE spectrum.
The lepto-hadronic model\cite{Reynoso:2010pp} fits to the low energy
$\gamma$-ray spectrum by Fermi/LAT and HESS low state but not the flaring state.
The spine-sheath model by
Tavecchio and Ghiselline\cite{Tavecchio:2008be} has difficulties to achieve a harder spectrum
in the VHE range due to strong absorption of the TeV photons from
interactions with the optical-infra red (IR) photons from the
spine. In the jet-in-jet model of Giannios et al.\cite{Giannios:2009pi} minijets are formed
within the jet due to flow instabilities and these minijets move
relativistically with respect to the main jet flow. The interaction of
the daughter jets with the main jet are responsible for the production
of VHE gamma rays. While the minijets are aligned with our line of
sight, the VHE gamma rays are beamed with large Doppler factor.
This scenario can explain the 2010 flare but does not provide a
quantitive prediction of the light curve of the flare.
Similarly, the magnetosphere model \cite{Rieger:2007tt,Levinson:2010fc,Rieger:2011ch} 
 can explain the hard TeV
spectrum but in this case also there is no detailed quantitive predication for the
VHE light curve. Similarly in the work by Cui et al.\cite{Cui:2011aa} can
explain the VHE gamma ray flare in an external inverse Compton model
with a very wide jet to have a Doppler boosting.
Borkov et al.\cite{Barkov:2012hd} have proposed a scenario where a red giant star with a loosely
bound envelope of mass $\sim 10^{29}$ {\rm g} interacts with the base
of the M87 jet. The VHE
emission is produced near the SMBH due to the interaction of the
cosmic ray protons emerging from the jet with the disrupted dense
cloud of the red giant through proton-proton interaction.
This model reproduced well  the light curve and the energy
spectrum of the April 2010 flare. But how universal is this scenario ? 
Can we be sure that VHE 
flaring in other AGNs/Blazars happen only due to the interaction of
respective jet with the intervening cloud from a foreign object  ?  If
so, can we be able to explain the multiple episodic flaring of these objects. Possibly,
 the 2010 flaring of M87 might be due to the jet cloud
interaction, but it is unlikely to be universal. Limitations of the
above discussed models and the non-universality of the jet cloud
scenario can be overcome in an alternative scenario  presented and applied to the orphan
TeV flaring of the blazar 1ES 1959+650\cite{Sahu:2013ixa}. A similar
scenario is also invoked to explain the multi-TeV emission from
Centaurus A\cite{Sahu:2012wv}. In this mechanism, the low  energy tail
of the SSC photons or the SSC peak serves as the target for the Fermi-accelerated high
energy protons to produce the pions
through delta resonance and their subsequent decay to high energy
photons and neutrinos. This scenario neither needs any intervening foreign object
nor any special jet cloud geometry\cite{Bottcher:2004qs} to produce the high energy
photons. 

The plan of the paper is as follows: In Sec.\ref{themodel} we review
in detail the flaring model and the kinematical condition for the
production of $\Delta$-resonance. The results are discussed in
Sec.\ref{result} and we briefly conclude in  Sec.\ref{final}.

\section{The Flaring Model}
\label{themodel}

In a recent paper Sahu et al.\cite{Sahu:2013ixa} have explained  the orphan TeV flare
of 4th June, 2002, from the blazar 1ES1959+650 through hadronic model. In this work they use
 the standard interpretation of the leptonic model to explain both, low and high 
energy peaks, by synchrotron and SSC photons respectively as in the
case of any other AGNs and Blazars. Thereafter, they propose that the
low energy tail of the SSC photons in the blazar jet
serve as the target for the Fermi-accelerated high energy protons of
energy $\le 100$ TeV, within the jet to produce TeV photons through
the decay of neutral pions from the delta resonance. 
 This model explains very nicely the observed TeV flux from the orphan flare. Also
it is interesting to note that, this scenario is self sufficient and 
does not need any external medium for the production of  gamma
rays. As discussed, the flaring 
occurs within a  compact and confined
volume of radius $R'_f$ inside the blob of radius $R'_b$ ($R'_f <
R'_b$) which is shown in Figure 1 of ref.\cite{Sahu:2013ixa} . 
Both the internal and the external jets are moving with the
same bulk Lorentz factor 
$\Gamma$ and the Doppler factor ${\cal  D}$ as the blob. In normal
situation within the jet, we consider the injected spectrum of the
Fermi accelerated charged particles having a power-law spectrum 
$dN/dE\propto
E^{-\alpha}$ with the power index $\alpha \ge 2$. But in the flaring
region we assumed that the Fermi accelerated charged particles have a
power-law with an an exponential cut-off spectra\cite{Aharonian:2003be,Sahu:2013ixa}, it is given as
\be
\frac{dN_p}{dE_p}\propto  E_p^{-\alpha} e^{-E_p/E_{p,c}},
\label{powerlaw}
\ee
where the high energy proton has the cut-off energy $E_{p,c}$ and
again the spectral index has the restriction $\alpha > 2$. 
Also probably due to the 
copious annihilation of electron positron
pairs, splitting of photons in the magnetic field, enhance IC
photons and Poynting flux dominated flow
from the magnetic  reconnection in the strongly
magnetized plasma around the base of the
jet\cite{Giannios:2009kh,Giannios:2009pi}, the comoving photon density
$n'_{\gamma, f}$ (flaring) in the flaring region is much higher than
the rest of the blob $n'_{\gamma}$ (non-flaring) i.e.
${n'_{\gamma, f}(\epsilon_\gamma)}\gg
{n'_{\gamma}(\epsilon_\gamma)}$. Here we assume that
the ratio of photon densities at two different
background energies $\epsilon_{\gamma_1} $  and $\epsilon_{\gamma_2} $
in flaring and non-flaring states remains the same, that is
\be
\frac{n'_{\gamma, f}(\epsilon_{\gamma_1})}
{n'_{\gamma, f}(\epsilon_{\gamma_2})}=\frac{n'_\gamma(\epsilon_{\gamma_1})}
{n'_\gamma(\epsilon_{\gamma_2})}.
\label{densityratio}
\ee

In general, in the leptonic one-zone synchrotron and SSC jet  model
the emitting region is a blob
with comoving radius $R'_b$  moving with a velocity
$\beta_c$ corresponding to a bulk Lorentz
factor $\Gamma$ and seen at
an angle $\theta_{ob}$ by an observer which results with a Doppler
factor ${\cal D}=\Gamma^{-1} (1-\beta_c \cos\theta_{ob})^{-1}$.  The emitting region is filled with an
isotropic electron population and a randomly oriented magnetic field
$B'$. The electrons have a power-law spectrum.
The energy spectrum of the Fermi-accelerated protons in the blazar jet
is also assumed to be of power-law. Due to high radiative losses, electron
acceleration is limited. On the other hand, protons and heavy nuclei can
reach UHE through the same acceleration mechanism.

Due to photohadronic interaction in the jet, the pions are produced
through the intermediate $\Delta$-resonance and is given by
\be
p+\gamma \rightarrow \Delta^+\rightarrow  
 \left\{ 
\begin{array}{l l}
 p\,\pi^0, & \quad \text {fraction~ 2/3}\\
  n\,\pi^+ , 
& \quad  \text {fraction~ 1/3}\\
\end{array} \right. ,
\label{decaymode}
\ee
which has a cross section $\sigma_{\Delta}\sim 5\times 10^{-28}\,
{\rm cm}^2$. Subsequently, the charged and neutral pions  will
decay through
$\pi^+\rightarrow e^+{\nu}_e\nu_{\mu}{\bar\nu}_{\mu}$ and
$\pi^0\rightarrow\gamma\gamma$ respectively. The produced neutrinos and photons are
in the GeV-TeV range energy. For the production of
$\Delta$-resonance, the kinematical condition is 
\be
E'_p \epsilon'_\gamma= \frac{(m^2_{\Delta}-m^2_p)} {2 (1-\beta_p
  \cos\theta)},
\label{kinecondi}
\ee
where $E'_p$ and $\epsilon'_\gamma$ are respectively 
the proton and the background  photon energies in the comoving frame
of the jet.  We define the quantities with a prime  in the comoving
frame and without prime in the observer frame. 
For high energy protons we assume
$\beta_p\simeq 1$. Since in the comoving frame the protons collide
with the SSC photons
from all directions, in our calculation we consider an average value $(1-\cos\theta) \sim 1$ 
($\theta$ in the range of 0 and $\pi$). 
Going from comoving frame to observer frame, the proton and photons
energies can be written as
\be
E_p=\frac{\Gamma E'_p}{(1+z)}, 
\ee
\be
\epsilon_\gamma = \frac{{\cal D} \epsilon'_\gamma}{(1+z)}
\ee
respectively and the kinematical condition given in
Eq.(\ref{kinecondi})  can be written in the observer frame as
\be
E_p \epsilon_\gamma \simeq 0.32~ \frac{\Gamma {\cal D}}{(1+z)^2} ~{\rm GeV}^2~.
\label{resonant1} 
\ee
In the jet comoving frame, each pion carries $\sim 0.2$ of the proton
energy while 50\% of the $\pi^0$ energy will be given to each
$\gamma$. So the relationship between high energy
$\gamma$-ray and the $E_p$ is $E_{\gamma}={\cal D} E_p/10$. From these relations we can express the
$\Delta$-resonance kinematical condition in terms of photon energies
(target photon energy $\epsilon_{\gamma}$ and the observed photon energy
$E_{\gamma}$) as
\be
E_\gamma \epsilon_\gamma \simeq 0.032~\frac{ {\cal D}^2}{(1+z)^2} ~{\rm GeV}^2.
\label{Eegamma} 
\ee
The optical depth  to produce  the $\Delta$-resonance is given as
\be
\tau_{p\gamma}=n'_{\gamma, f}\sigma_{\Delta} {\cal R}'_f.
\label{optdep}
\ee 
The comoving photon number density within the confined volume can be
given in terms of the luminosity $L_{\gamma}$ as 
\be
n'_{\gamma, f} = \eta \frac{L_\gamma}{{\cal D}^{2+\kappa}} \frac{(1+z)}{4\pi
{\cal R'}^2_f \,\epsilon_\gamma},
\ee
with $\kappa \sim (0-1)$
(depending on whether the jet is continuous or discrete) and $\eta
\sim 1$ . Here in this work we consider $\kappa=0$. For $\kappa=1$,
the photon density will be reduced by a of factor ${\cal D}^{-1}$ in
the discrete jet as compared to continuous one.
The relationship between observed $\gamma$-ray flux $F_{\gamma}$, high
energy proton flux and the background SSC photon density in the
flaring region is given
as\cite{Sahu:2013ixa}
\bary
F_{\gamma}(E_{\gamma}) &\equiv& E^2_{\gamma} \frac{dN(E_\gamma)}{dE_\gamma} \nonumber\\
&\propto & E^2_p \frac{dN(E_p)}{dE_p} n'_{\gamma, f}(\epsilon_\gamma).
\eary
Then the observed high energy $\gamma$-ray flux at two different
energies will scale as 
\be
\frac{F_\gamma(E_{\gamma_1})}{F_\gamma(E_{\gamma_2})} 
= \frac{n'_\gamma(\epsilon_{\gamma_1})}
{n'_\gamma(\epsilon_{\gamma_2})}
\left(\frac{E_{\gamma_1}}{E_{\gamma_2}}\right)^{-\alpha+2}
e^{-(E_{\gamma_1}-E_{\gamma_2})/E_c}~,
\label{spectrum}
\ee
where $E_{\gamma_{1,2}}$ are two different $\gamma$-ray energies and
correspondingly the proton energies are $E_{p_{1,2}}$. In the above
equation (\ref{spectrum}) we have used the relations in
Eq.(\ref{densityratio}) and
\be
\frac{E_{p_1}}{E_{p_2}}=\frac{E_{\gamma_1}}{E_{\gamma_2}},.
\ee
From an observed flux in a given
energy, we can calculate the fluxes at other
energies by using Eq.(\ref{spectrum}).

The optical depth $\tau_{p\gamma}$ implies that out of
$\tau^{-1}_{p\gamma}$ many protons, one will interact with the SSC
background photons to produce $\Delta$-resonance. In this case the
fluxes of the TeV photons and the
Fermi accelerated high energy protons $F_p$, are related through
\be
F_p(E_p)=7.5\times \frac{ F_{\gamma}(E_{\gamma})}{\tau_{p\gamma}(E_p)},
\label{protonflux}
\ee
Like photons, the proton fluxes at different energies $E_{p_1}$ and
$E_{p_2}$, scale as
\be
\frac{F_p(E_{p_1})}{F_p(E_{p_2})}=\left ( \frac{E_{p_1}}{E_{p_2}}
\right)^{-\alpha+2} e^{-(E_{p_1}-E_{p_2})/E_{p,c}}.
\ee
From this relation we can calculate the proton fluxes at different energies.

\begin{figure}[t!]
{\centering
\resizebox*{0.5\textwidth}{0.4\textheight}
{\includegraphics{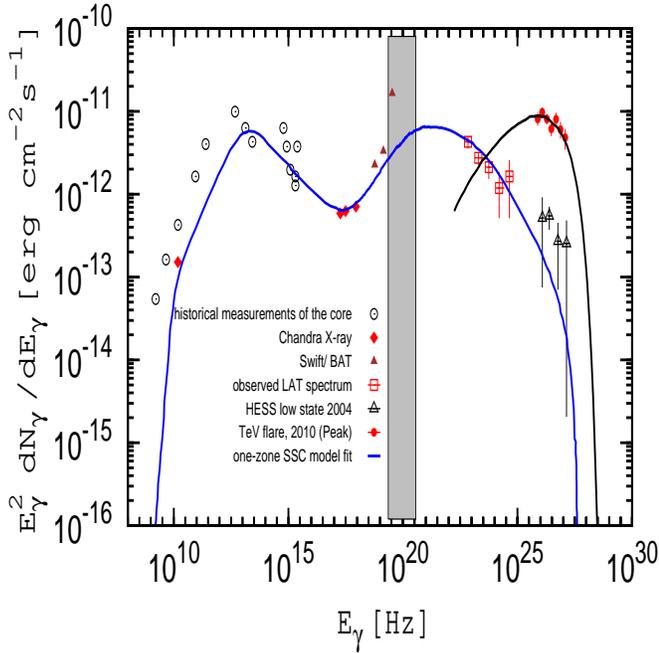}}
\par}
\caption{The SED of M87 is shown in all the energy band. The peak flux
  of the flare of April 2010 is only shown here. The hadronic model
  fit to the 2010 data is shown as continuous line to the extreme
  right. The shaded region is the energy range of SSC photons where the
  Fermi-accelerated protons are collided to produce the $\Delta$-resonance.
}
\label{m87sed}
\end{figure}
\begin{figure}[t!]
{\centering
\resizebox*{0.5\textwidth}{0.4\textheight}
{\includegraphics{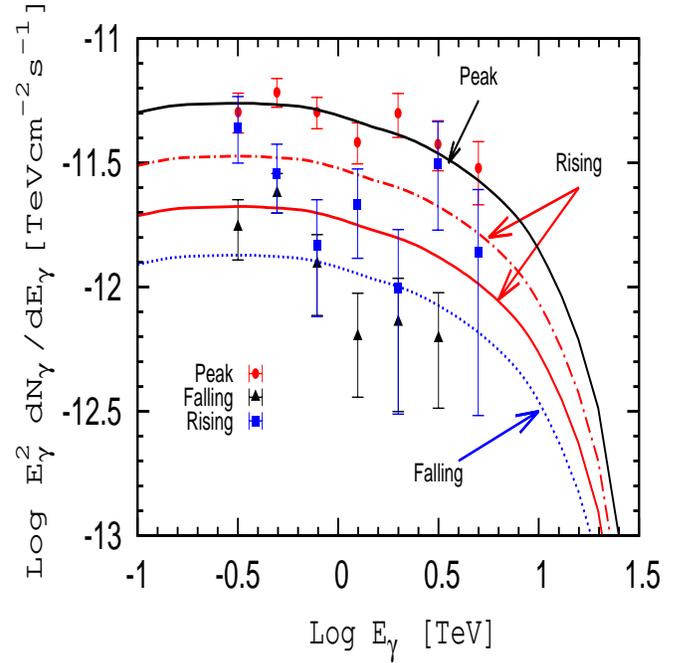}}
\par}
\caption{The rising, the peak and the falling parts of the TeV flare
  are fitted with the power-law exponential SED. For all these we use
  the same spectral index $\alpha=2.83$ and $E_c\simeq 12$ TeV. The rising
  part is fitted with two different normalized flux.
}
\label{risepeakfall}
\end{figure}

\begin{figure}[t!]
{\centering
\resizebox*{0.5\textwidth}{0.4\textheight}
{\includegraphics{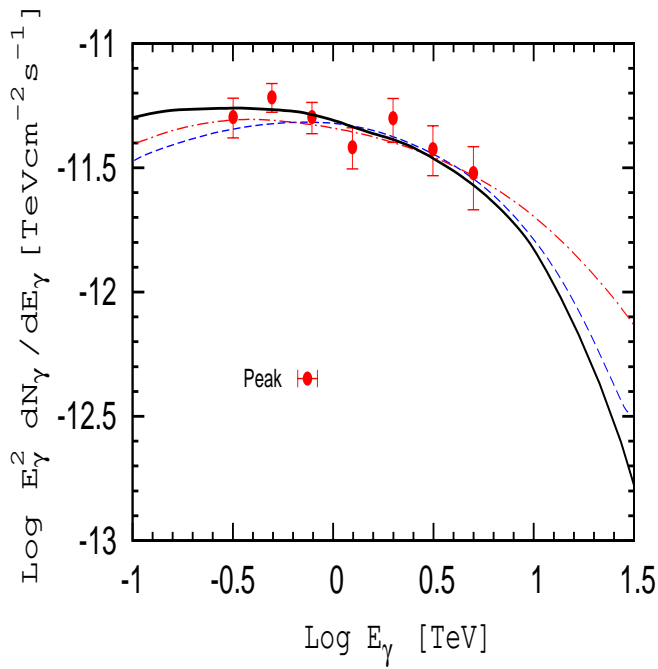}}
\par}
\caption{ The dashed-dotted and dashed lines are the  SED calculated
  in the jet cloud interaction
  hadronic model of ref.\cite{Barkov:2012hd} for two different
  injection spectra and  fitted to the peak
  SED of the flare. The continuous curve is the hadronic model fit.
}
\label{peakonly}
\end{figure}

\begin{figure}[t!]
{\centering
\resizebox*{0.5\textwidth}{0.4\textheight}
{\includegraphics{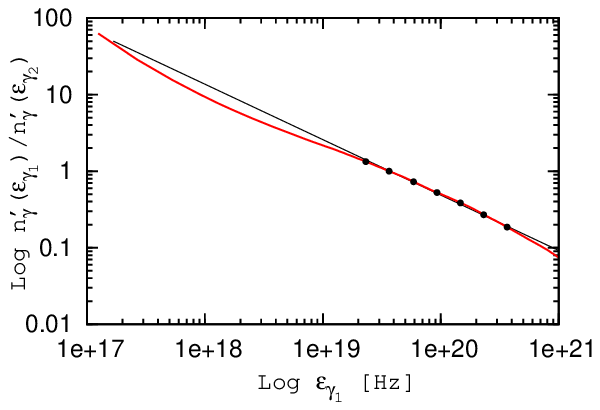}}
\par}
\caption{
The ratio of photon densities
$n'_{\gamma}(\epsilon_{\gamma_1})/n'_{\gamma}(\epsilon_{\gamma_2})$
 for a given value  $n'_{\gamma}(\epsilon_{\gamma_2}=3.7\times
10^{19}\, {\rm Hz})$ is plotted 
as a function of SSC photon energy. The points are the density ratios
for observed data points. The curve is fitted with a straight line.
}
\label{ng1ng2}
\end{figure}
\section{Results}
\label{result}

With the homogeneous leptonic one-zone synchrotron and SSC  jet model\cite{Finke:2008pe}
the SED is fitted assuming the viewing angle $10^{\rm o}$ and bulk
Lorentz factor $\Gamma=2.3$ which corresponds to a Doppler factor
${\cal D }=3.9$ which is shown in Figure 4 of Ref.\cite{Abdo:2009ta}.  
Based on the multi-band correlations
detected in the 2005 and 2008 flaring events of M87, the core and
the HST-1 are favored as the emitting regions. But during the VHE
flare of 2008 and 2010,  {\it Chandra} detected an enhanced x-ray flux from the
core region which are the two highest measurements since the start of
its observation in 2002. During these time HST-1 remained in a low
state\cite{Harris:2011zf}.  During the 2005 VHE flaring episode no enhanced x-ray
emission from the core was detected. On the other hand, at that time,
HST-1 was more than 30 times brighter than the core region in x-rays
leading to uncertainty in the flux estimation of the core\cite{Harris:2009wn} .
The coincidence in x-ray and VHE emission as well as  the observed
timescales of short variability ($\sim$ day) at VHE/x-ray suggests that
the size of the emitting region is compact  lead to believe that
the 2010 VHE flare probably originates in the innermost region of the
jet. So here we assume that the flaring
occurs within the confined volume of radius $R'_f = 5\times 10^{15}$
cm which is in the core region. For the fit to the multi-wavelength SED in ref.\cite{Abdo:2009ta}, 
the source radius is taken to be $R'_{b}=1.4\times 10^{16}$ cm which is
consistent with the few day timescale variability in TeV and the
magnetic field is $B=55$ mG. Also this is consistent with $R'_{f} <
R'_{b}$. In this work we use the SED and parameters of the one-zone
synchrotron model given in ref.\cite{Abdo:2009ta}.

During the flaring in April 2010, the high energy $\gamma$-ray flux was observed in the energy
range $0.3 ~ {\rm TeV } \lesssim E_{\gamma} \lesssim ~5$ TeV. Also the
flaring had distinct rise time and fall time of the spectra. The
rising, the peak and the falling of the flux are fitted with power-law with different flux
normalization and the spectral index $\alpha$\cite{Aliu:2011xm}.  In the 
hadronic model, the above $E_{\gamma}$ range corresponds to the proton
energy in the range $1.9 ~{\rm TeV} \lesssim E_{p} \lesssim 30~ {\rm
 TeV}$. Protons in this energy range will
collide with the background photons in the energy range $1.5~{\rm MeV}
(3.7\times 10^{20}\, {\rm Hz} )\gtrsim \epsilon_{\gamma} \gtrsim 0.1\,
{\rm MeV} (2.3\times 10^{19}\, {\rm Hz})$ to produce
$\Delta$-resonance and subsequent decay of it will produce both
$\gamma$-rays and neutrinos through neutral and charged pion decay respectively. We
can observe that the above $\epsilon_{\gamma}$
lie in the rising part of the SSC photons shown as shaded region in
Figure \ref{m87sed}.  The
number density of these photons are also calculated which lie in the
range $72\,{\rm cm^{-3}} \lesssim n'_{\gamma}\lesssim\, 516 \,{\rm
  cm^{-3}}$.

As discussed in the ref.\cite{Sahu:2013ixa}
for the calculation of the TeV flux, first we take into account one of
the observed flaring fluxes with its corresponding energy for
normalization e.g. 
$F_{\gamma}(E_{\gamma_2}=3.18\,{\rm TeV})\simeq 3.8\times 10^{-12}\,{\rm TeV}\,{\rm  cm}^{-2}\,{\rm
  s}^{-1}$
and  $n'_{\gamma}(\epsilon_{\gamma_2}=0.15
{\rm MeV})\simeq 387\, {\rm cm^{-3}}$ and using it calculate the flux for other energies
with the Eq.(\ref{spectrum}). This we have done for
different observed fluxes for a better fit.
The spectral index
$\alpha$ and the cut-off energy $E_c$  are the free parameters in the
model and  the best fit is obtained for $\alpha=2.83$ and $E_c\simeq 12$
TeV. The $\gamma$-ray cut-off
energy of $\sim$12 TeV corresponds to $E_{p,c}\simeq 71$ TeV and above the cut-off
energy the flux decreases rapidly which is clearly shown in Figure \ref{m87sed}. With the same $\alpha=2.83$ and
$E_c\simeq 12$ TeV  but different normalized flux we
fitted the rising, the peak and the falling flux which are shown in
Figure \ref{risepeakfall}. The rising flux is fitted with two different normalized
flux to have a better picture. In Figure \ref{peakonly}, the fitting
of the peak flux in our model is compared with the 
cloud-jet interaction model\cite{Barkov:2012hd}. 
It is interesting to note that the spectral index $\alpha$ fitted to the
TeV flaring SEDs of M87 and the blazar 1 ES 1959+650 have the same
value 2.83 which probably hints for a common
mechanism of particle acceleration during the flaring\cite{Sahu:2013ixa}.

We have also plotted the ratio of photon densities
$n'_{\gamma}(\epsilon_{\gamma_1})/n'_{\gamma}(\epsilon_{\gamma_2})$ of
Eq. (\ref{spectrum} ) for a given value  $n'_{\gamma}(\epsilon_{\gamma_2}=3.7\times
10^{19}\, {\rm Hz})\simeq 386\,{\rm cm^{-3}}$
in Figure \ref{ng1ng2} as a function of SSC photon energy. It shows
that the density ratio is almost a linear function of energy. We have
specifically chosen the energy range in the vicinity of the shaded
region of Figure \ref{m87sed} which is responsible for the TeV  spectra. 
The standard power-law fitting with an exponential cut-off to the SED\cite{Aharonian:2003be} is
expressed as
\be
F=F_0 \left (\frac{E}{\rm TeV}\right )^{-\alpha+2} e^{-E/E_c},
\ee
 where $F_0$ is a constant. But here $F_0$ is replaced by energy
 dependent photon density of the background and due to this energy
 dependent coefficient, fitting to SED in this model is different from
 the standard one.

During the flaring period, not only protons but also electrons are
Fermi-accelerated in the inner jet with the same energy as the
protons. The $e^+$ produced during the $\pi^+$ decay has energy in the
range $0.095\,{\rm TeV} \lesssim E_{\gamma} \lesssim 1.5 \, {\rm
  TeV}$. These electrons and positrons will produce synchrotron
radiation in the jet magnetic field. While the Fermi-accelerated
electrons will emit synchrotron photons in the frequency band
$2.5\times 10^{18}\, {\rm Hz} \lesssim \epsilon_{\gamma} \lesssim
6.3\times 10^{20}\, {\rm Hz} $, the positrons will radiate in the
frequency band $6.3\times 10^{15}\, {\rm Hz} \lesssim \epsilon_{\gamma} \lesssim
1.6\times 10^{18}\, {\rm Hz} $. So the flaring in the TeV energy
should be accompanied by a simultaneous enhanced synchrotron emission in the frequency band
$6.3\times 10^{15}\, {\rm Hz} \lesssim \epsilon_{\gamma} \lesssim
6.3\times 10^{20}\, {\rm Hz} $.

\section{Conclusions}
\label{final}

The strongest TeV flaring of the radio galaxy M87 in April 2010 is
explained by assuming it to be due to the photohadronic interaction of the
Fermi-accelerated protons of energy $\lesssim 30$ TeV with the SSC
photons in the energy range $\sim (0.1 -1.5) MeV$. In this scenario
the proton spectrum is a power-law with an exponential cut-off.
For the fitting of the rising, the peak and the falling parts of the
TeV flare we use the same spectral index $\alpha=2.83$ and the
$\gamma-$ray cut-off energy $E_c\simeq 12$ TeV. Our results fit well to these
distinct phases of the flare.

\acknowledgements

E.P. is very much thankful to ICN, UNAM, Mexico City  for their kind hospitality
during his several visits. This work is partially supported by DGAPA-UNAM (Mexico) Project 
No. IN103812.

\end{document}